 \documentclass[
preprintnumbers,a4paper, amsmath, amssymb, floatfix,
nofootinbib, twocolumn, wrapfloat byrevtex]{revtex4}

 \usepackage{amsmath,mathrsfs,graphicx,epstopdf,placeins,float}
 \usepackage{booktabs}
 \usepackage{multirow}
 \usepackage{pstricks}

 \newcommand{\beq}[1]{\begin{equation}\label{#1}}
 \newcommand{\eeq}{\end{equation}}
 \newcommand{\bea}[1]{\begin{eqnarray}\label{#1}}
 \newcommand{\eea}{\end{eqnarray}}

\begin{document} 

 \title{Morris-Thorne wormhole in the vector-tensor theories with Abelian gauge symmetry breaking }
 
 \author{Ai-chen Li $^{a,b}$}
\email{lac@emails.bjut.edu.cn , aichenli@icc.ub.edu}
 \author{Xin-Fei Li$^{c}$}
\email{xfli@gxust.edu.cn}
\affiliation{\it ${}^a$ Institut de Ci\`encies del Cosmos, Universitat de Barcelona, Mart\'i i Franqu\`es 1, 08028 Barcelona, Spain} 
\affiliation{\it ${}^b$ Departament de F\'isica Qu\`antica i Astrof\'isica, Facultat de F\'isica, Universitat de Barcelona, Mart\'i i Franqu\`es 1, 08028 Barcelona, Spain} 
 \affiliation{\it ${}^c$ School of Science, Guangxi University of Science and Technology, Guangxi, 545026, China}
 \begin{abstract}
 We construct an asymptotically flat Morris-Thorne wormhole solution supported by anisotropic matter fluid and a vector field which is coupled to gravity in a nonminimal way with broken Abelian gauge symmetry. In this paper, a specific shape function is considered. We find that the ansatz of vector field plays a significant  role in determining the spacetime geometry of the wormhole. If there exists the electrostatic potential only, the redshift function could be considered as a constant value, implying the vanishing tidal force. However, when the vector potential in radial-direction is involved, the r-component of extended Maxwell equations at the wormhole's throat is invalid. To solve this issue, a thin shell is introduced near the throat, dividing the spacetime into two parts. Furthermore, it is proved that the spacetime geometry of wormhole could be smooth at junction position if the expressions of redshift function and vector potential are given appropriately. Finally, the energy conditions and the volume integral quantifier are explored.
 
 \end{abstract}
\maketitle

\section{Introduction}

In general relativity, wormholes are interesting spacetime structures bridging two asymptotic regions located in one universe or multiverse \cite{VisserBook}, which are solutions of the Einstein field equations. Actually, the original motivation of introducing wormholes is to replace the singularity of Schwarzschild black hole by a 'tunnel structure' geometry, namely, 'Einstein-Rosen bridge' (ERB) \cite{Flamm, Einstein:1935tc, Ellis}.

In 1980s, Morris and Thorne introduced the traversable wormholes, which increase the possibility of spacetime traveling \cite{Morris:1988cz}. Furthermore, Visser has adopted the cut-and-paste method to construct the traversable wormholes known for thin-shell wormholes (TSW) \cite{Visser:1989kh, Visser:1989kg}, which are stable under the linear perturbation \cite{Poisson:1995sv}. Recently, the traversable wormholes have received broad attention in several aspects, such as the stability analysis from new perturbative method \cite{Lobo:2003xd, Lemos:2008aj, Li:2018jxy}, the resolution to the horizon problem in cosmology \cite{Hochberg:1992du, Kim:2018aaw}, finding the wormhole solutions from the models beyond the Einstein gravity \cite{Bhawal:1992sz, Hochberg:1990is, Agnese:1995kd, Jusufi:2020yus, Huang:2020qmn,  Ibadov:2020btp}. According to physical grounds, all the matter in our universe should satisfy certain energy conditions. While, the traversable wormholes require the existence of exotic matters which suffer from the violation of classical energy conditions, like the Weak Energy Condition (WEC) \cite{Morris:1988tu}, the Null Energy Condition (NEC) \cite{Hochberg:1998ii} and the Strong energy condition (SEC) \cite{Hochberg:1998vm}. Thus, it is a valuable research topic to find the traversable wormholes which conform to some classical energy conditions, especially the WEC and NEC. In \cite{Kanti:2011jz, Rosa:2018jwp}, the traversable wormholes have been constructed in some modified gravity model which belongs to the low-energy effective theory of string, without needing any form of exotic matter. In addition, inspired by the ER=EPR conjecture \cite{Maldacena:2013xja, Maldacena:2017axo}, \cite{Cariglia:2018rhw, Blazquez-Salcedo:2020czn} find the traversable-wormhole solutions in Einstein gravity with entangled fermions.

In recent decades, the scalar-tensor theories (ST) play a vital role in giving the alternative explanation on the origin of inflation and dark energy \cite{Nojiri:2003vn}, and pleanty of insightful physics have been explored in ST \cite{Zloshchastiev:2004ny, Winstanley:2005fu, Zeng:2009fp, Herdeiro:2014goa, Brihaye:2014nba}. However, few attentions have been drawn on the vector-tensor theories (VT). Actually, some interesting cosmological phenomenology, such as driving the accelerated expansion of universe at late time \cite{Tasinato:2014eka}, explaining the cosmological constant problem \cite{Tasinato:2014mia} and cosmic inflation \cite{Golovnev:2008cf, Koivisto:2008xf,DeFelice:2016yws,Emami:2016ldl}, could also be achieved by coupling the vector field to the gravity with broken Abelian gauge symmetry. In order to develop the phenomenology of VT theories in more areas of physics, our purpose in this paper is to consider the static and spherically symmetric Morris-Thorne wormholes for a type of non-minimally coupled vector-tensor theory with Abelian gauge symmetry breaking in four-dimensional spacetime \cite{Heisenberg:2014rta, Chagoya:2016aar} (hereafter, this model is called VTAB for brief). As shown by \cite{Chagoya:2016aar}, the hairy black hole solutions in VTAB should take inclusion of a nontrivial configuration of vector potential $A_r (r)$ besides the electrostatic potential $A_0 (r)$. Thus, it is worthwhile to consider the effects of $A_r(r)$ on traversable wormhole solutions. In particular, we will study the possibility that the matter satisfies the NEC everywhere, from the throat to infinity when some physical parameters are appropriately chosen.

This work is structured as follows. In Sec.\ref{AdSProWHSol}, at first, we briefly introduce a type of vector-tensor theory with broken Abelian gauge symmetry in four-dimensional spacetime. Then, the Einstein field equations and extended Maxwell equations are given in metric ansatz of Morris-Thorne wormhole geometry. In Sec.\ref{AdSProWHSolTriChi}, when vector potential $\chi$ vanishes, an asymptotically flat wormhole solution is presented. Besides, the NEC, the WEC and volume integral quantifier are  analyzed. The effects of vector potential on wormhole geometry have been explored in Sec.\ref{AdSProWHSolNonTriChi}. In particular, in order to avoid the divergence of redshift effects at throat of wormhole, we construct a pair of special piecewise functions for $\chi$ and $\Phi$, respectively. Meanwhile, these piecewise functions make the spacetime continuous at joining position. Finally, conclusions and discussions are presented in Sec.\ref{ConDiss}.

\section{General setup of wormhole in Vector-Tensor theory with Abelian symmetry breaking  \label{AdSProWHSol}}

Firstly, let us briefly review on the vector-tensor theory in four-dimensional spacetime  as in \cite{Chagoya:2016aar}. Its action is set as
\begin{align}
\nonumber
&S=\frac{1}{2\kappa^{2}}\int\sqrt{-g}~d^{4}x\big\{ R-2\Lambda-\frac{1}{4}F^{2}\\
\label{EinProBHaction}
&\quad\quad\quad\quad\quad\quad\quad\quad\quad+\beta G_{\mu\nu}A^{\mu}A^{\nu}+\mathcal{L}_{\text{fluid}}\big\}
\end{align}
in which  $G_{\mu\nu}$ is the standard Einstein tensor, $\beta$ is the physical constant  measuring the strength of nonminimal coupling between vector field and Einstein tensor, which indicates that the $U(1)$ symmetry is broken in presence of this nonminimal coupling term. In general case, the exotic matter  violating classical energy conditions is needed is introduced in order to keep up the geometry of traversable wormholes. Thus, we involve an extra matter content which has the form of an anisotropic fluid. From $\eqref{EinProBHaction}$, the Einstein field equation is given by
\begin{align}
\label{EinFieEqua}
& R_{\mu\nu}-\frac{1}{2}g_{\mu\nu}R+\Lambda g_{\mu\nu}=T_{\mu\nu}^{(0)}+T_{\mu\nu}^{(A)}\\
\nonumber
\\
\label{EMTofExoMatt}
&T_{\nu}^{(0)\mu}=\text{diag}\{-\rho,\mathcal{P}_{r},\mathcal{P}_{t},\mathcal{P}_{t}\}\\
\label{EMTofVecFie}
&T_{\mu\nu}^{(A)}=\frac{1}{2}\big(g^{\beta\alpha}F_{\nu\beta}F_{\mu\alpha}-\frac{1}{4}g_{\mu\nu}F^{2}\big)+\beta Z_{\mu\nu}\\
\nonumber
&Z_{\mu\nu}=\frac{1}{2}A^{2}R_{\mu\nu}+\frac{1}{2}RA_{\mu}A_{\nu}-2A^{\alpha}R_{\alpha(\mu}A_{\nu)}\\
\nonumber
&\quad\quad-\frac{1}{2}\nabla_{\mu}\nabla_{\nu}A^{2}+\nabla_{\alpha}\nabla_{(\mu}(A_{\nu) }A^{\alpha})-\frac{1}{2}\nabla^{\alpha}\nabla_{\alpha}(A_{\mu}A_{\nu})\\
\nonumber
&\quad\quad+\frac{1}{2}g_{\mu\nu}\big(G_{\alpha\beta}A^{\alpha}A^{\beta}+\nabla^{\alpha}\nabla_{\alpha}A^{2}-\nabla_{\alpha}\nabla_{\beta}(A^{\alpha}A^{\beta})\big)
\end{align}
in which the energy-momentum tensor $T^{(0)}_{\mu\nu}$ is derived from the $\mathcal{L}_{fluid}$, $\rho$ is the energy density, while $\mathcal{P}_r$ and $\mathcal{P}_t$ represent the pressures in the radial direction and transverse direction, respectively. The equation of motion for vector field, namely the extended Maxwell equations, reads
\begin{align}
\label{procaequa}
\nabla_{\mu}F^{\mu\nu}+2\beta A_{\mu}G^{\mu\nu}=0.
\end{align}
We assume $A_\mu$ has the following ansatz
\begin{align}
\label{ElecPotenAnsatz}
A_{\mu}dx^{\mu}=a(r)dt+\chi(r)dr,
\end{align}

We consider the static and spherically symmetric metric in four-dimensional spacetime, with the following ansatz as in \cite{Chagoya:2016aar},
\begin{align}
\label{MTWormAnsatz}
&ds^2=-e^{\Phi(r)}dt^2 +\frac{dr^2}{1-\frac{b(r)}{r}}+r^2 (d\theta^2+\sin^2\theta d\phi^2).
\end{align}
where $\Phi(r)$ is the redshift function for an infalling observer,  and $b(r)$ represents the spatial shape function of the wormhole geometry. In order to avoid the presence of an event horizon, the redshift function $\Phi(r)$ should be finite everywhere. Two asymptotic spacetime regions are connected by the throat of wormhole which is located at the minimum radial coordinates $r_0$, with the condition that $b(r_0)=r_0$. Moreover, the flaring-out condition of wormhole geometry requires the shape function $b(r)$ to satisfy
\begin{align}
\label{flareout}
\frac{b-b^{\prime}r}{2b^{2}}>0
\end{align}
which reduces to $b^\prime (r_0)<1$ at the throat of wormhole. Besides, to avoid the coordinate singularity in region $r>r_0$, the  restriction is given by
\begin{align}
\label{AvoidHoriz}
&1-\frac{b(r)}{r}>0
\end{align}

After substituting ansatz $\eqref{ElecPotenAnsatz}$ and $\eqref{MTWormAnsatz}$ into Einstein field equations $\eqref{EinFieEqua}$ and the extended Maxwell equations $\eqref{procaequa}$, the following indepentent differential equations are given
\begin{widetext}
	\begin{align}
	\label{Vecr}
	&0=\frac{\chi}{r^{2}}\big(\Phi^{\prime}r(r-b)-b\big)\\
	\label{Vect}
	&0=2r(b-r)a^{\prime\prime}+\big(4b+(b^{\prime}-4)r\big)a^{\prime}+4\beta b^{\prime}a\\
	\label{Eintt}
	&\rho=\frac{e^{-\Phi}}{4r^{4}}\bigg(r^{2}\big((b-r)ra^{\prime2}-2\beta a^{2}b^{\prime}\big)-2e^{\Phi}\big(\beta\chi^{2}(b-r)(3rb^{\prime}-2b-2r)+4\beta r\chi\chi^{\prime}(b-r)^{2}-2r^{2}b^{\prime}\big)\bigg)\\
	\label{Einrr}
	&\mathcal{P}_{r}=\frac{e^{-\Phi}}{4r^{3}}\bigg(8\beta raa^{\prime}(r-b)+(r-b)\big(4\beta e^{\Phi}\chi^{2}+r^{2}a^{\prime2}\big)-4\beta a^{2}b\bigg)\\
	\nonumber
	&\mathcal{P}_{t}=\frac{1}{4}\big(\frac{r\mathcal{P}_{r}+b\rho}{r-b}+3\mathcal{P}_{r}+2r\mathcal{P}_{r}^{\prime}\big)=\frac{e^{-\Phi}}{8r^{4}(r-b)}\bigg(\frac{e^{\Phi}b^{3}\chi}{2}(\chi-2r\chi^{\prime})+r^{2}b\big(\frac{a^{2}}{4}(4+3b^{\prime}+4r\Phi^{\prime})\\
	\nonumber
	&\quad~+e^{\Phi}(2b^{\prime}+\chi^{2}+\frac{7}{4}\chi^{2}b^{\prime}-5r\chi\chi^{\prime})+r^{2}a^{\prime}(2ra^{\prime}\Phi^{\prime}+a^{\prime}b^{\prime}-7a^{\prime}-4ra^{\prime\prime})+ra(2a^{\prime}b^{\prime}+4ra^{\prime}\Phi^{\prime}-3a^{\prime}-4ra^{\prime\prime})\big)\\
	\nonumber
	&\quad~+rb^{2}\big(-\frac{a^{2}}{2}(3+2r\Phi^{\prime})-\frac{1}{4}e^{\Phi}\chi(3b^{\prime}\chi+6\chi-16r\chi^{\prime})+ra(a^{\prime}+2ra^{\prime\prime}-2ra^{\prime}\Phi^{\prime})+r^{2}a^{\prime}(3a^{\prime}+2ra^{\prime\prime}-ra^{\prime}\Phi^{\prime})\big)\\
	\label{Einthetatheta}
	&\quad~+r^{3}\big(e^{\Phi}\chi(2r\chi^{\prime}-b^{\prime}\chi)-a^{2}b^{\prime}+2ra(ra^{\prime\prime}+a^{\prime}-a^{\prime}b^{\prime}-ra^{\prime}\Phi^{\prime})+r^{2}a^{\prime}(2ra^{\prime\prime}+4a^{\prime}-a^{\prime}b^{\prime}-ra^{\prime}\Phi^{\prime})\big)\bigg)
	\end{align}
\end{widetext}
In this work, we  restrict our attention to the asymptotically flat solutions, thus $\Lambda=0$  in $\eqref{Eintt}-\eqref{Einthetatheta}$. Besides, as indicated in \cite{Chagoya:2016aar}, the asymptotically flat solutions with a nontrival configuration of $A_\mu$ could be obtained only if $\beta=\frac{1}{4}$. Hereafter, we will set $\beta=\frac{1}{4}$ throughout the whole paper.

\section{Asymptotically flat solutions $\uppercase\expandafter{\romannumeral1}$: a specific shape function with $\chi=0$ \label{AdSProWHSolTriChi} }

\begin{figure}[ht]
	\begin{center}
		\includegraphics[scale=0.336]{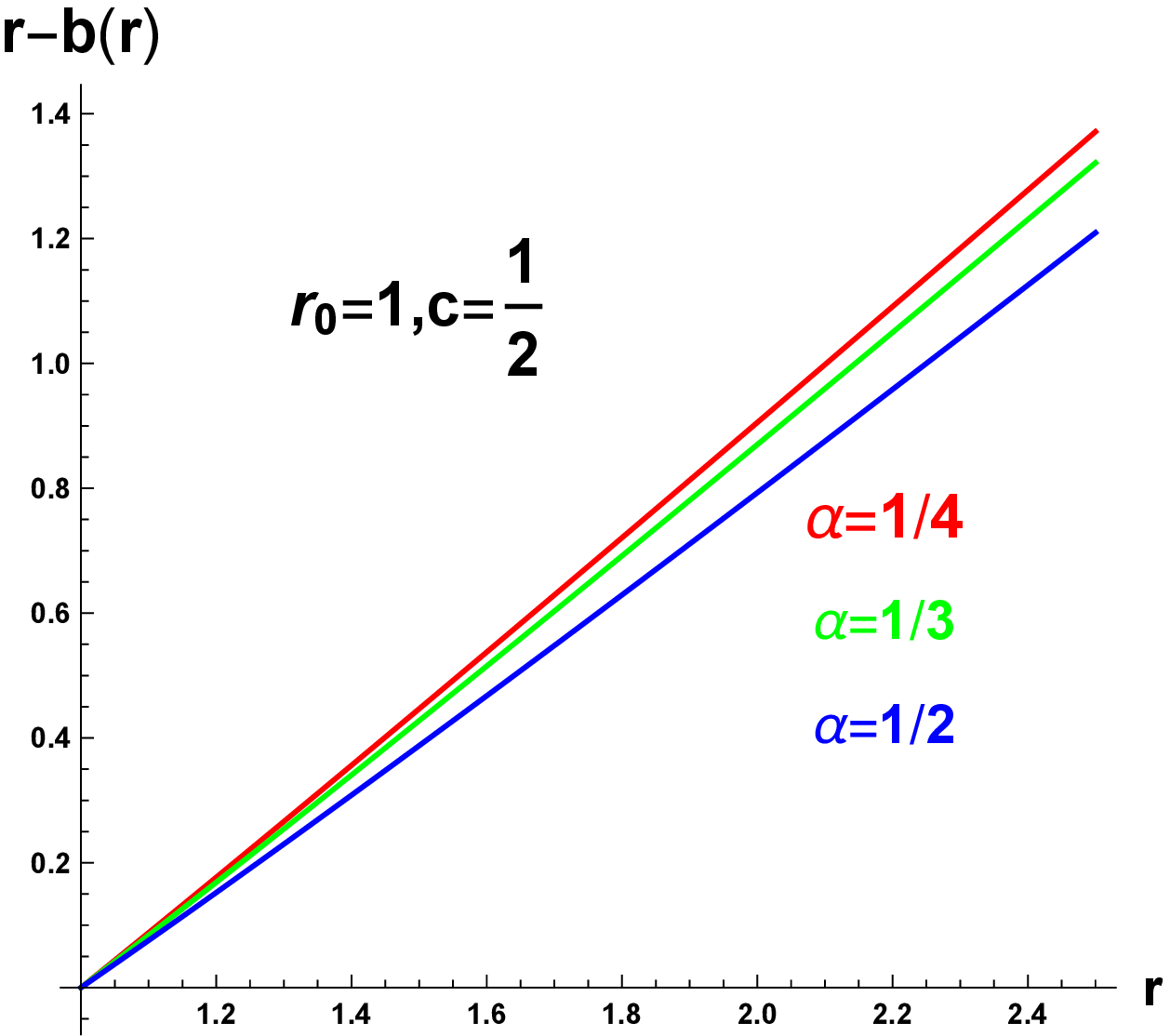}
		\includegraphics[scale=0.336]{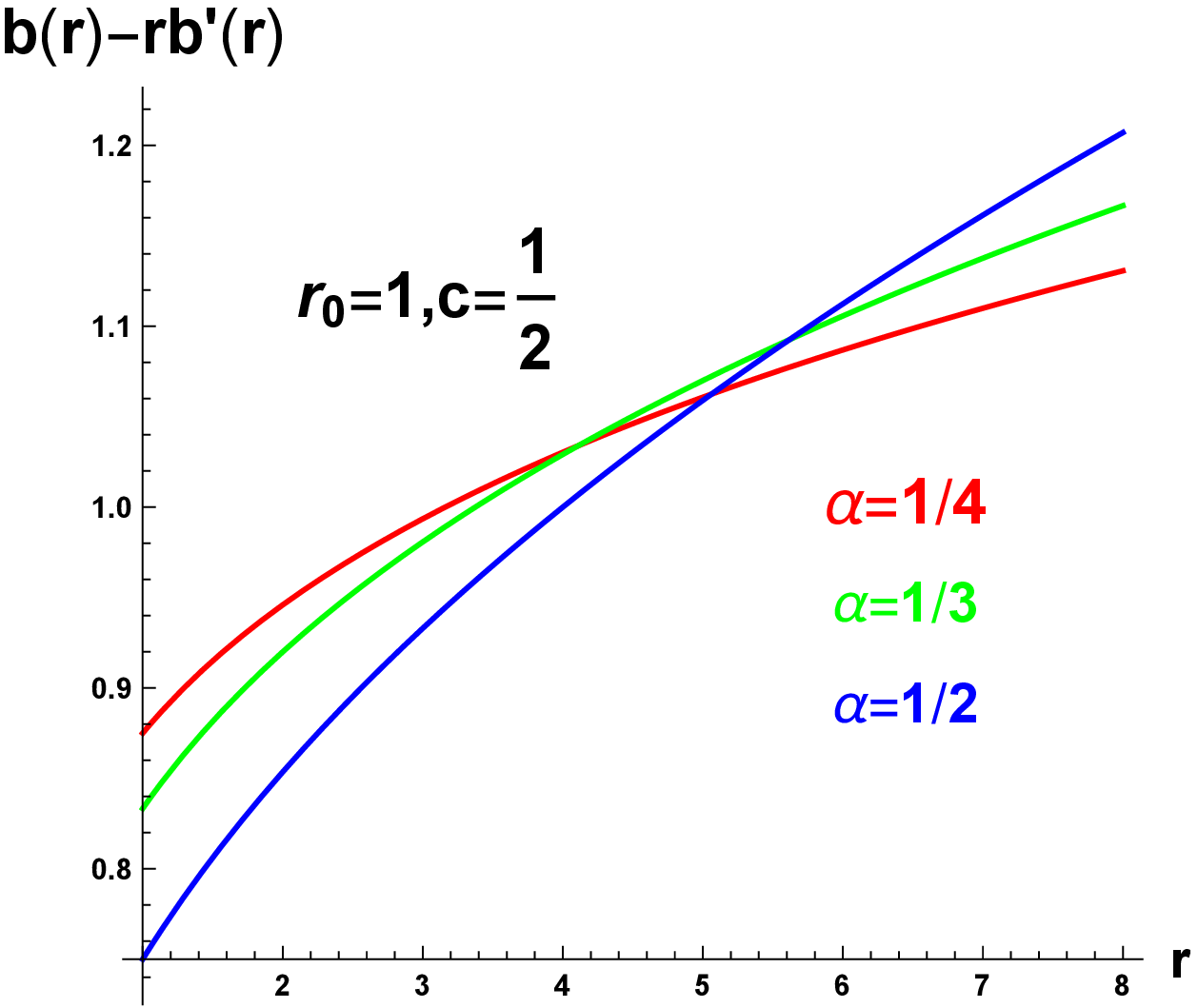}
		\caption{(color online). The aim of this figure is to show that the conditions $\eqref{flareout}$ and $\eqref{AvoidHoriz}$ for the specific shape function $\eqref{SpeShape1}$ is not difficult to be satisfied by choosing some representative parameters according to the inequalities $\eqref{FlaCon1}$ and $\eqref{FlaCon2}$.}
		\label{ConWHgeo}
	\end{center}
\end{figure}

\begin{figure}[ht]
	\begin{center}
		\includegraphics[scale=0.335]{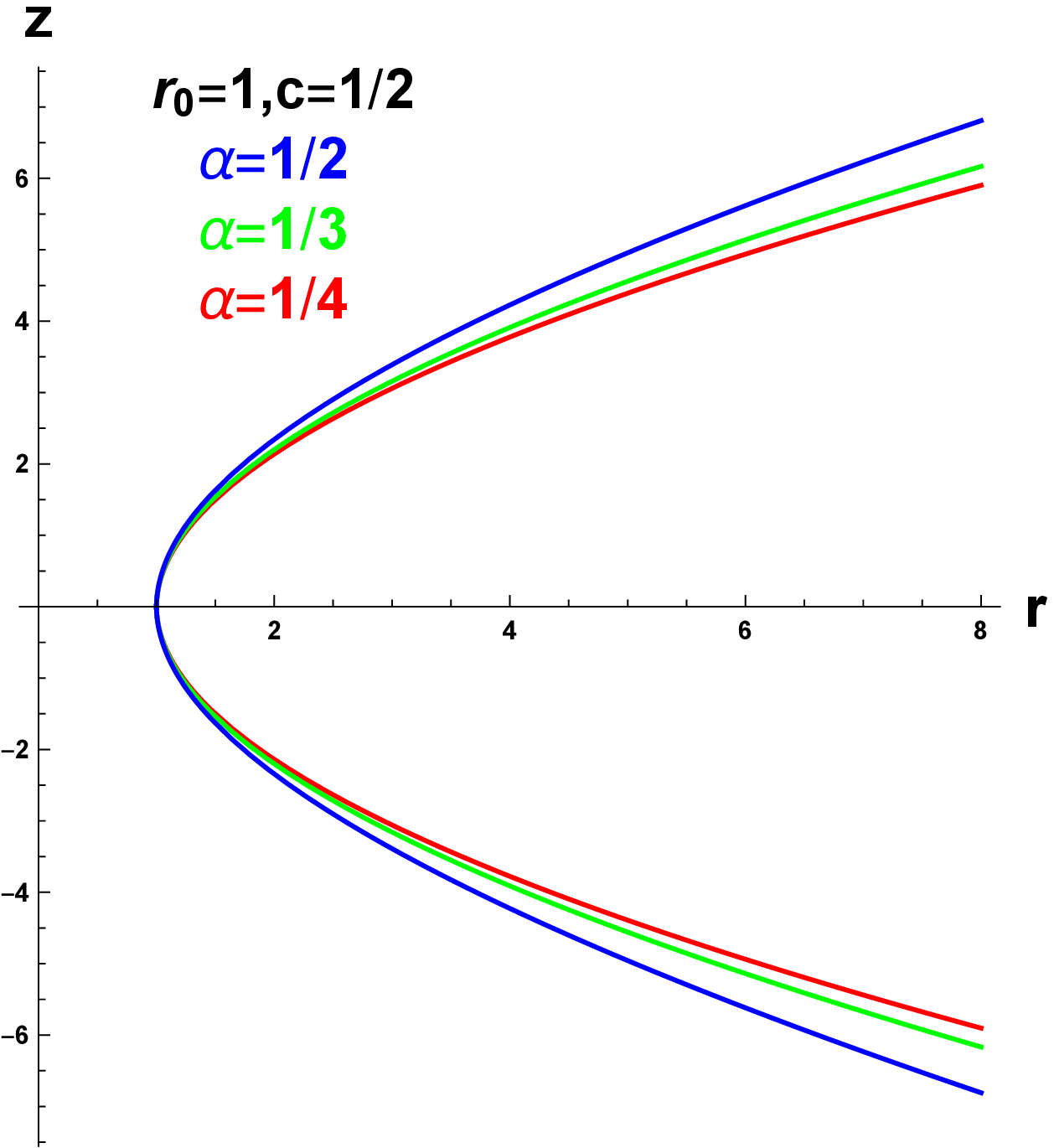}
		\includegraphics[scale=0.335]{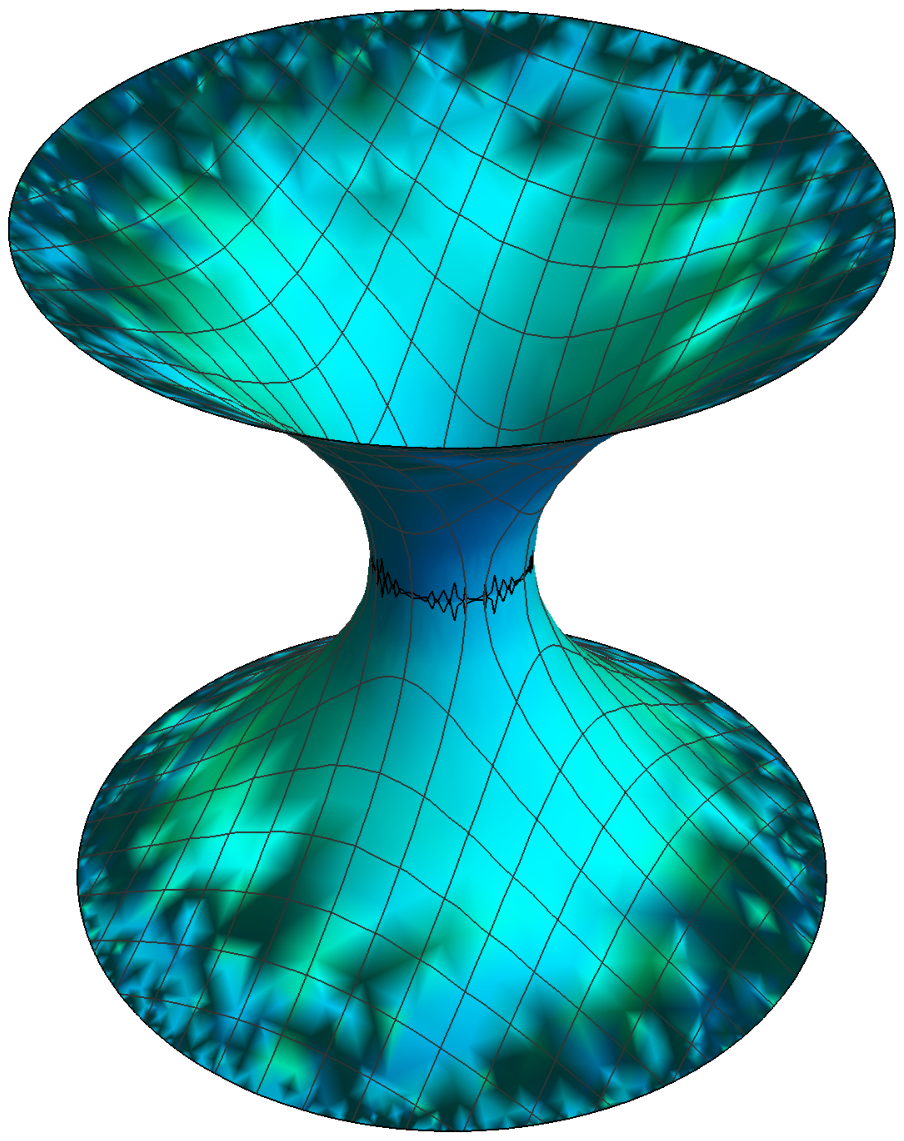}
		\caption{(color online). The embedding diagram of wormhole geometry along the equatorial plane $\theta=\pi/2$ at a fixed time. The left panel is the configuration of $z(r)$ solved from the \eqref{CacZr}. While the right one is the corresponding 3-dimensional diagram emerged from $z(r)$ by sweeping through a $2\pi$ rotation around the $z$-axis.}
		\label{EmbedSurface}
	\end{center}
\end{figure}

Following the sprit of \cite{Lobo:2012qq}, we will consider a type of specific shape function in this work
\begin{align}
\label{SpeShape1}
&\frac{b(r)}{r_{0}}=c(\frac{r}{r_{0}})^{\alpha}+1-c
\end{align}
Since our attention is concentrated on the asymptotically flat solutions, i.e. $\frac{b(r)}{r}\big\vert_{r\to\infty}=0$, the condition $\alpha<1$ is imposed. Besides, the conditions $\eqref{flareout}$ and $\eqref{AvoidHoriz}$ imply the following inequalities
\begin{align}
\label{FlaCon1}
&\quad \quad\quad \quad\quad ~ c\alpha<1\\
\label{FlaCon2}
&c\alpha(\frac{r}{r_{0}})^{\alpha}<1-c+c(\frac{r}{r_{0}})^{\alpha}<\frac{r}{r_{0}}
\end{align}
In Fig.\ref{ConWHgeo}, some typical $c, \alpha$ parameters are given to make $\eqref{FlaCon1}-\eqref{FlaCon2}$ hold.
Substituting $\eqref{SpeShape1}$ into $\eqref{Vect}$, one yields to
\begin{align}
\label{elecPotenSha1}
&a=\frac{Q}{r},
\end{align}
which is the standard coulomb potential. According to the works \cite{Chagoya:2016aar, Li:2020kcw}, it is necessary to indicate the following facts about the charge $Q$. Since the $U(1)$ symmetry is broken due to the nonminimal coupling term $\beta G_{\mu \nu} A^\mu A^\nu$. Thus the $Q$ is not a conserved quantity any more. And here we have to consider this quantity in the grand canonical ensemble, in which the system can exchange the charge particles with the exterior and the numbers of charges is variable (the conjugate variable of charge, i.e. the chemical potential $\mu$, is constant). It is straightforward to observe that the equation $\eqref{Vecr}$ will be trivial when $\chi=0$. And then, we consider a constant redshift function, i.e. $\Phi(r)=\Phi_0$, in order to simplify the problem. For this case, the metric becomes
\begin{align}
\nonumber
&ds^{2}=-dt^{2}+\frac{dr^{2}}{1-c(\frac{r}{r_{0}})^{\alpha-1}-(1-c)\frac{r_{0}}{r}}\\
\label{Spec1Metric}
&~~~~~+r^{2}(d\theta^{2}+\sin^{2}\theta d\phi^{2})
\end{align}
Note that here the factor $e^{-2\Phi_{0}}$ is absorbed into $dt^{2}$ through the redefinition of time coordinate. From $\eqref{Spec1Metric}$, one could analyze the embedding diagram of wormhole geometry into the Euclidean space. Without loss of generality,  an equatorial slice $\theta=\pi/2$ at a fixed time $t=const$ are considered. Then the metric $\eqref{Spec1Metric}$ reduces to
\begin{align}
\label{ReduMetric}
&ds^2=\frac{dr^{2}}{1-c(\frac{r}{r_{0}})^{\alpha-1}-(1-c)\frac{r_{0}}{r}}+r^{2} d\phi^{2},
\end{align}
which could be embedded into a 3-dimensional Euclidean space with cylindrical symmetry, namely
\begin{align}
\label{3DEuclidean}
&ds^2_E=dz^2+dr^2+r^2d\phi^2.
\end{align}
By matching $\eqref{ReduMetric}$ with $\eqref{3DEuclidean}$, the embedded surface $z(r)$ is obtained as
\begin{align}
\label{CacZr}
&\frac{dz}{dr}=\pm\sqrt{\frac{1}{1-c(\frac{r}{r_{0}})^{\alpha-1}-(1-c)\frac{r_{0}}{r}}-1}
\end{align}
We could evaluate the integral $\eqref{CacZr}$ numerically for specific parameters, and the corresponding
profile of $z(r)$ is shown in Fig.\ref{EmbedSurface}.

\begin{figure}[ht]
	\begin{center}
		\includegraphics[scale=0.3]{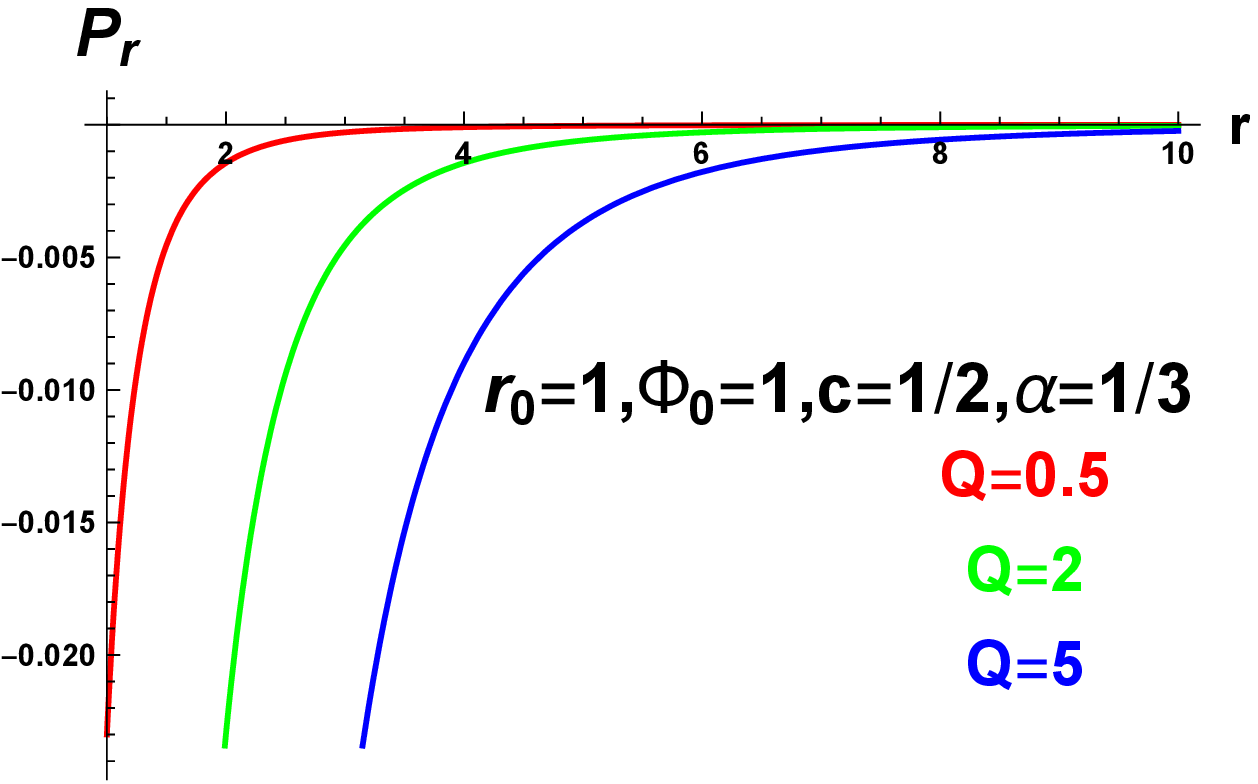}
		\includegraphics[scale=0.3]{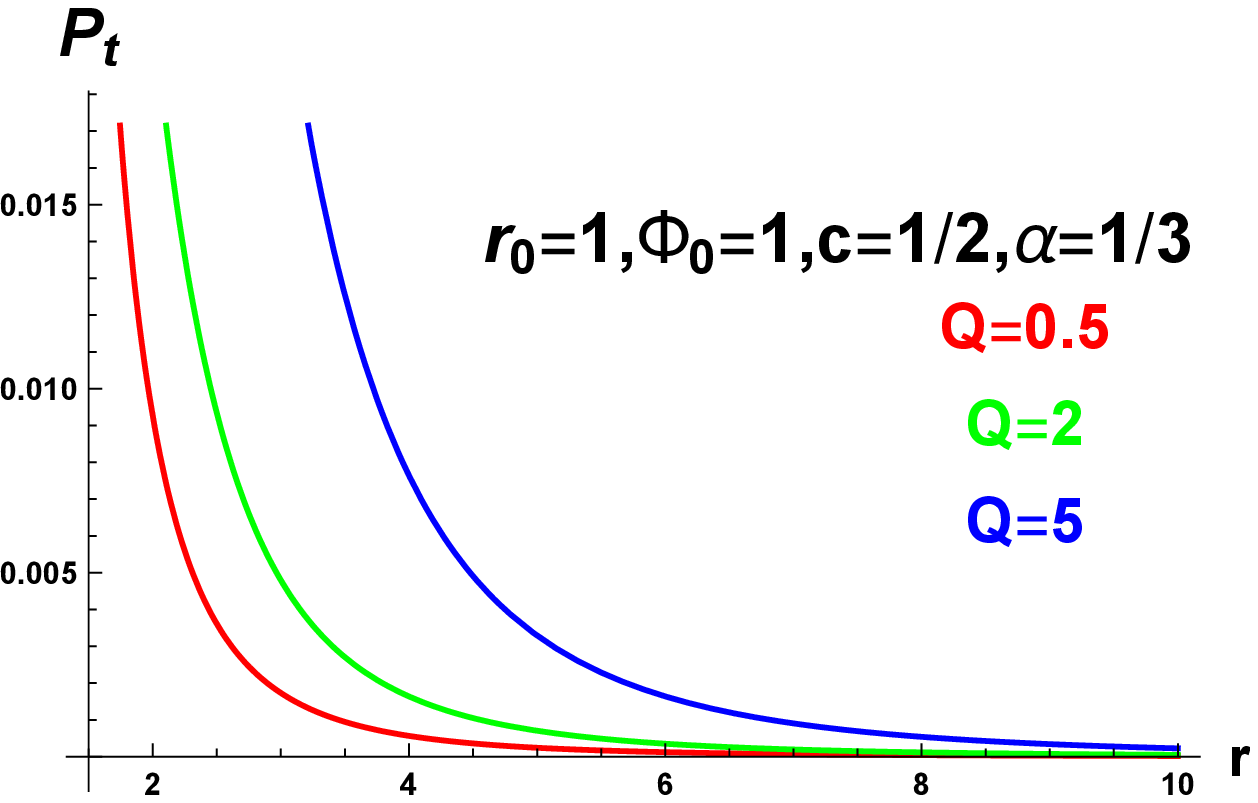}\\
		\includegraphics[scale=0.3]{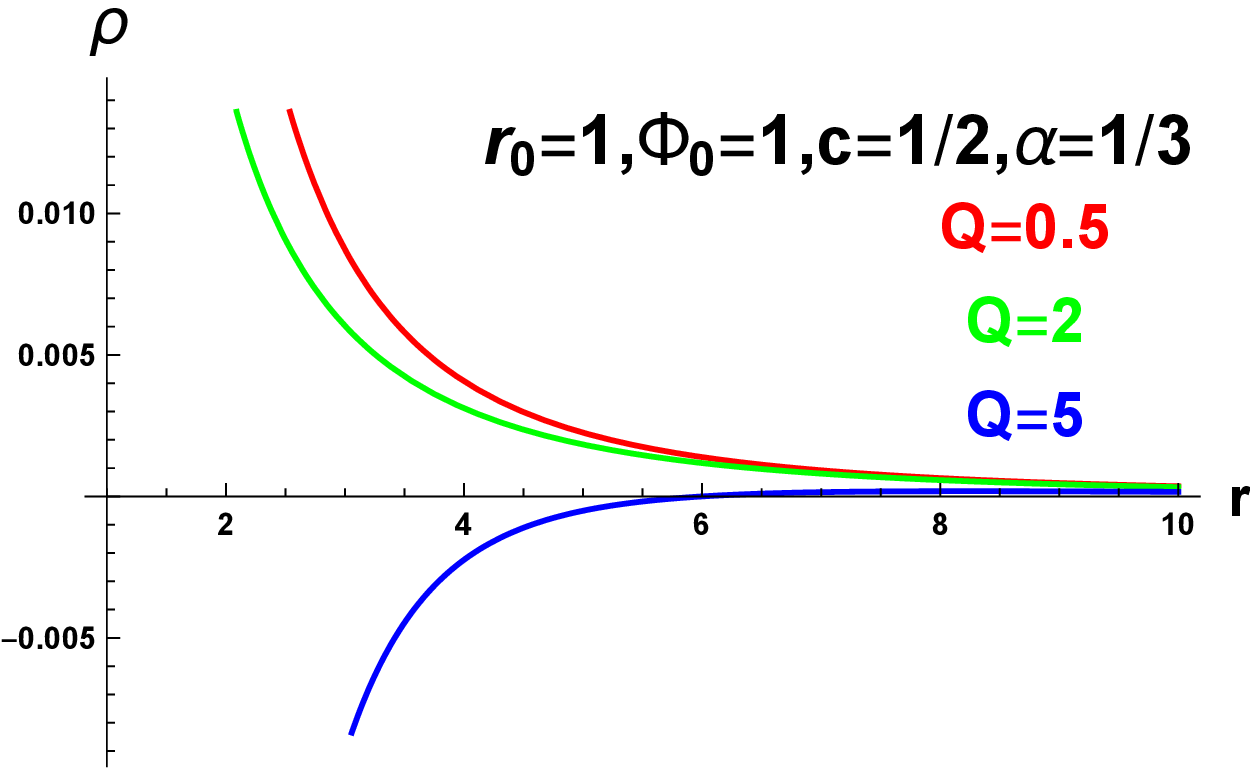}
		\includegraphics[scale=0.3]{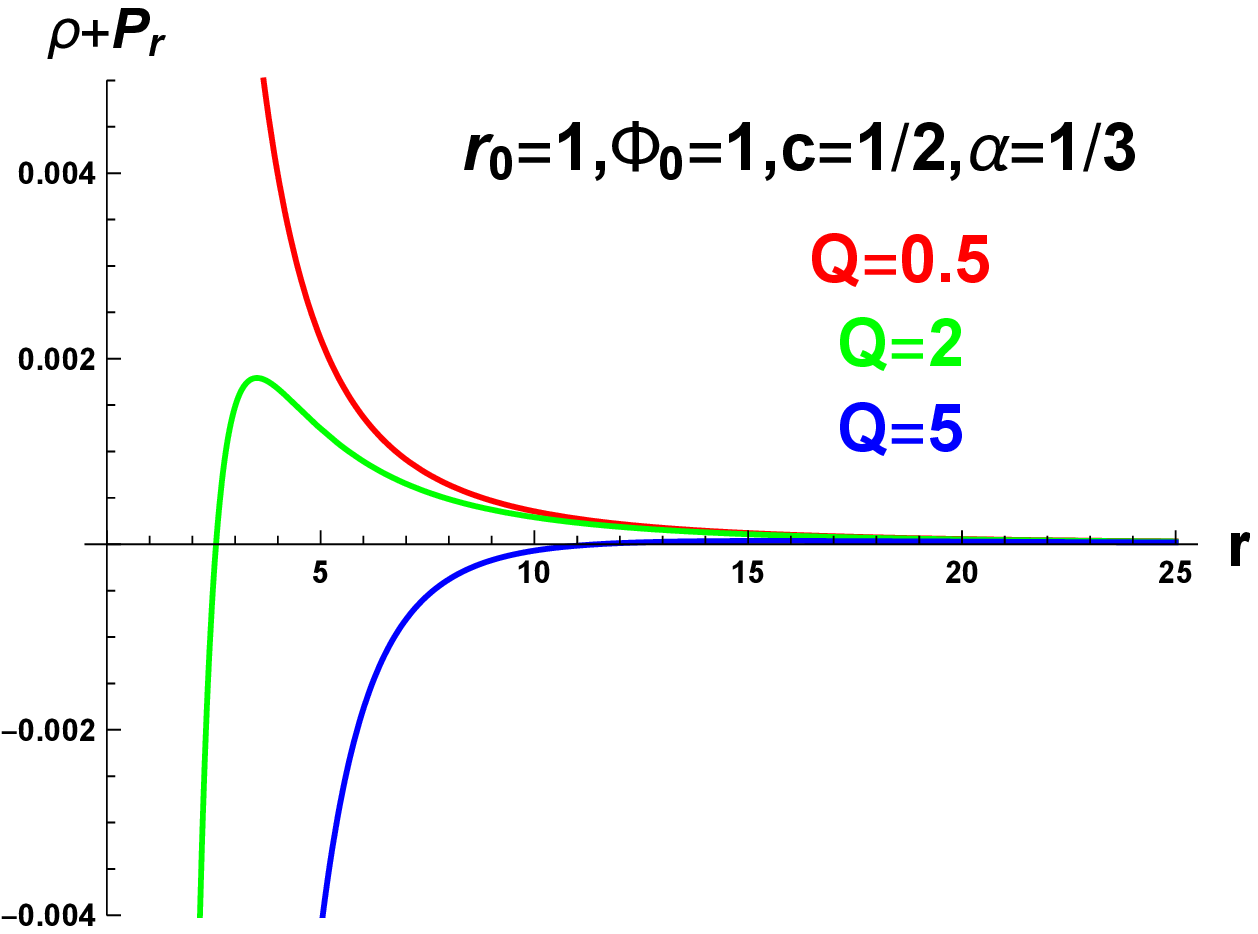}
		\caption{(color online). Plot the variation of $\rho,~P_r,~P_t,~\rho+P_r$} with respect to the $r$ at fixed $r_0,~\phi_0,~c,~\alpha$ in different $Q$.
		\label{EMvsQspec1}
	\end{center}
\end{figure}

Plugging $\eqref{SpeShape1}$ and $\eqref{elecPotenSha1}$ into $\eqref{Eintt}-\eqref{Einthetatheta}$, respectively,
\begin{align}
\nonumber
&\hspace{-3mm}\rho=\frac{e^{-\Phi_{0}}}{8r^{5}}\bigg(2Q^{2}(r_{0}-r)+2Q^{2}cr_{0}\big((\frac{r}{r_{0}})^{\alpha}-1\big)\\
\label{rhochie0}
&\hspace{-3mm}\quad-\alpha Q^{2}cr_{0}(\frac{r}{r_{0}})^{\alpha}+8\alpha cr_{0}^{3}e^{\Phi_{0}}(\frac{r}{r_{0}})^{\alpha+2}\bigg)\\
\label{Prchie0}
&\hspace{-3mm}\mathcal{P}_{r}=-\frac{e^{-\Phi_{0}}}{4r^{4}}Q^{2}\\
\nonumber
&\hspace{-3mm}\mathcal{P}_{t}=\frac{e^{-\Phi_{0}}}{32r^{5}\big(r-r_{0}+cr_{0}(1-(\frac{r}{r_{0}})^{\alpha})\big)}\bigg\{8Q^{2}r^{2}\\
\nonumber
&\hspace{-3mm}\quad-4cQ^{2}r_{0}^{2}+12Q^{2}rr_{0}(c-1)+2Q^{2}r_{0}^{2}(c^{2}+1)\\
\nonumber
&\hspace{-3mm}\quad+cr_{0}(\frac{r}{r_{0}})^{\alpha}\bigg(Q^{2}\big((\alpha-4)(c-1)r_{0}-(\alpha-2)cr_{0}(\frac{r}{r_{0}})^{\alpha}\big)\\
\label{Ptchie0}
&\hspace{-3mm}\quad-12r+8\alpha e^{\Phi_{0}}r^{2}r_{0}\big(1+c((\frac{r}{r_{0}})^{\alpha}-1)\big)\bigg)\bigg\}
\end{align}
From the expressions $\eqref{rhochie0}-\eqref{Ptchie0}$, the variation of $\rho, P_r, P_t$ with respect to  $r$ in some representative parameters are displayed in Fig.\ref{EMvsQspec1}.
It is worthwhile noting that the energy density $\rho$ will change from positive values to the negative ones as one increases the $Q$. In particular, let us consider the null energy condition (NEC) and weak energy condition (WEC) respectively for this wormhole solution. It is well known that the WEC is defined by $T_{\mu\nu}U^\mu U^\nu\geq 0$, i.e. $\rho \geq 0~\&~\rho(r)+P_r(r)\geq 0$, in which the $U^\mu$ should be a timelike vector. Meanwhile, the NEC satisfies $T_{\mu\nu}K^\mu K^\nu\geq 0$, i.e. $\rho(r)+P_r(r)\geq 0$, with $K^\mu$ being a null vector. From $\rho$-$r$ and $(\rho+P_r)$-$r$ curves in Fig.\ref{EMvsQspec1}, we see that the WEC and NEC hold in case of small $Q$, while both them are broken as $Q$ increases. In other words, it means that the wormhole could exist without introducing the exotic matter when $Q$ is small.

In case of the large $Q$, the total amount of exotic matter could be evaluated by "volume integral quantifier" \cite{Visser:2003yf, Kar:2004hc, Nandi:2004ku}, which is defined as
\begin{align}
\label{TotaMountExo}
I_V=2 \int ^{r_c} _{r_0} (\rho+\mathcal{P}_r) r^2 \sin\theta dr d\theta d\phi
\end{align}
where $r_c$ is the radius beyond which  $\rho+P_r$ has a positive value. One can check that the value of $\rho+P_r$ is finite at $r=r_0$. Thus, when $Q$ is large, the wormhole solution could be constructed with small quantities of exotic matter, which keeps up the flaring-out geometry near the throat.

\section{Asymptotically flat solutions $\uppercase\expandafter{\romannumeral2}$: a specific shape function with $\chi\ne 0$\label{AdSProWHSolNonTriChi}}

In case of $\chi\ne0$, eq. $\eqref{Vecr}$ is a non-trival equation. It implies the following equation
\begin{align}
\label{DeterPhi}
&\Phi^{\prime}r(r-b)-b=0
\end{align}
However, the $\eqref{DeterPhi}$ is invalid at the wormhole's throat $r=r_0$. Thus, for making $\eqref{Vecr}$ hold in all ranges of $[r_0,\infty)$, we assume  $\chi(r)$ with the form that
\begin{align}
\label{pisefunctionchi}
&\chi=\begin{cases}
\begin{array}{c}
0~,~(r_{0}\leq r<r_{1})\\
\chi_{+}(r)~,~(r\geq r_{1})
\end{array}\end{cases}
\end{align}
in which $r_1$ should satisfy $r_1>r_0$ and $\chi_+$ is an undetermined function. Combining $\eqref{Vecr}$ with $\eqref{pisefunctionchi}$ that
\begin{align}
\label{pisefunctionPhi}
&\hspace{-2.5mm}\Phi(r)=\begin{cases}
\begin{array}{c}
\hspace{-25mm}\Phi_{-}(r)~,~(r_{0}\leq r<r_{1})\\
\Phi_{+}(r)=\int_{r}^{\infty}\frac{b(r^{\prime})}{r^{\prime}\big(r^{\prime}-b(r^{\prime})\big)}dr^{\prime}~,~(r\geq r_{1})
\end{array}\end{cases}		
\end{align}
Here the boundary condition $\Phi_+(r)\big\vert_{r\to\infty}=0$ is imposed to guarantee that the spacetime is asymptotically flat at infinity.

\begin{figure}[ht]
	\begin{center}
		\includegraphics[scale=0.5]{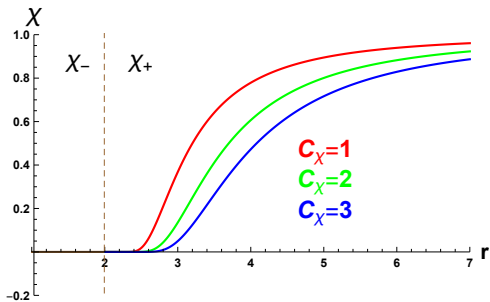}
		\caption{(color online). The shape of function $\chi(r)$ are plotted  in different $C_\chi$.}
		\label{chitest}
	\end{center}
\end{figure}

For the undetermined function $\chi$, we expect it to be continuous at least in first derivative at junction position $r=r_1$. Meanwhile, the value of $\chi$ should not  be divergent at infinity. Thus, throughout this section, we consider wormholes with the following $\chi$ function
\begin{align}
&\chi_+=\exp[-\frac{C_\chi}{(r-r_1)^2}]~,~C_\chi >0
\end{align}
In Fig.\ref{chitest}, we display  the  shape of $\chi$ function in radial direction $\eqref{pisefunctionchi}$.
In $\eqref{pisefunctionPhi}$, without loss of generality, we choose a specific value $\alpha=\frac{1}{2}$. And the $\Phi_+$ could be integrated analytically
\begin{align}
\label{AnaIntePhiPlus}
&\Phi_{+}(r)=\ln(\frac{r}{r_{0}})+\frac{1}{c-2}\ln\frac{(1-\sqrt{\frac{r}{r_{0}}})^{2}}{\big(1-c+\sqrt{\frac{r}{r_{0}}}\big)^{2(c-1)}}
\end{align}
Similarly, we also expect $\Phi_-$ to be continuous at least in first derivative at $r=r_1$. At the same time, the value of $\Phi_-$ is finite in regions $[r_0,r_1]$. Thus, a specific ansatz for $\Phi_-$ is chosen as
\begin{align}
\label{pisefunctionPhiMinu}
&\Phi_{-}(r)=\phi_{1}\sin\frac{r}{r_{0}}+\phi_{2}
\end{align}
Combine $\eqref{AnaIntePhiPlus}$ with conditions $\Phi_{-}(r_{1})=\Phi_{+}(r_{1})$ and $\Phi_{-}^{\prime}(r_{1})=\Phi_{+}^{\prime}(r_{1})$, the undetermined coefficients $\phi_1~,\phi_2$ are calculated as
\begin{align}
&\phi_{1}=\big(\frac{r_{0}}{r_{1}}-\frac{r_{0}}{r_{1}+(c-1-c\sqrt{\frac{r_{1}}{r_{0}}})r_{0}}\big)\sec\frac{r_{1}}{r_{0}}\\
\nonumber
&\phi_{2}=\ln(\frac{r_{1}}{r_{0}})+\frac{1}{c-2}\ln\frac{(1-\sqrt{\frac{r_{1}}{r_{0}}})^{2}}{\big(1-c+\sqrt{\frac{r_{1}}{r_{0}}}\big)^{2(c-1)}}\\
&\quad\quad-\big(\frac{r_{0}}{r_{1}}-\frac{r_{0}}{r_{1}+(c-1-c\sqrt{\frac{r_{1}}{r_{0}}})r_{0}}\big)\tan\frac{r_{1}}{r_{0}}
\end{align}
\begin{figure}[ht]
	\begin{center}
		\includegraphics[scale=0.5]{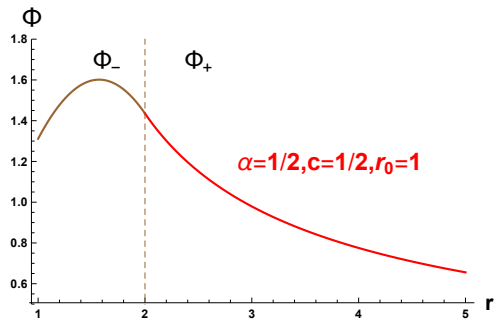}
		\caption{(color online). Plot the function of $\Phi(r)$ under the ansatzes $\eqref{pisefunctionPhi},\eqref{AnaIntePhiPlus},\eqref{pisefunctionPhiMinu}$.}
		\label{Phitest}
	\end{center}
\end{figure}
As shown by Fig.\ref{Phitest}, the redshift function $\Phi(r)$ is continuous at joining position.

Although the functions $\chi,~\Phi$ and their first derivatives are continuous at junction position $r=r_1$, there is no guarantee that the spacetime geometry is smooth  at the junction point. Thus, it is necessary to evaluate the junction condition \cite{Israel:1966rt} at hypersurface $r=r_1$. Specifically, we introduce a static hypersurface, namely the so-called thin-shell, at $r_1$, which connects the interior solution (denoted by subscript "-") and the exterior solution (denoted by subscript "+"). The intrinsic coordinates of the thin-shell are denoted as $x^a=(t,\theta,\phi)$. Therefore, the $4$-velocity of the thin-shell could be easily obtained as $u^\mu=(1,0,0,0)$, and the unit normal vector pointing into the hypersurface of the thin-shell is $n_\nu=(0,\frac{1}{\sqrt{1-\frac{b}{r}}},0,0)$. Furthermore, the vielbein $e^\mu_a$ is defined as $e^\mu_a=\frac{\partial x^\mu}{\partial x^a}$, which satisfies $n_\mu e^\mu_a=0$. Accordingly, the projection tensor is $\gamma_{\mu\nu}=g_{\mu\nu}-n_\mu n_\nu$, whose tangential components $\gamma_{ab}=e^\mu _a e^\nu _b\gamma_{\mu\nu}$ correspond to the induced metric of thin-shell, namely
\begin{align}
&ds^{2}=\gamma_{ab}dx^{a}dx^{b}=-e^{\Phi(r_{1})}dt^{2}+r_{1}^{2}d\theta^{2}+r_{1}^{2}\sin^{2}\theta
\end{align}
The junction condition for this vector-tensor theory has been derived by the work \cite{Li:2020kcw}, which is
\begin{align}
\nonumber
&\big\{\mathcal{K}_{ab}-\mathcal{K}\gamma_{ab}+\beta\big(\frac{1}{2}\gamma_{ab}A^{2}\mathcal{K}+\frac{1}{2}\gamma_{ab}n^{\rho}\nabla_{\rho}A^{2}\\
\nonumber
&-A^{2}\mathcal{K}_{ab}-\gamma_{ab}n_{\alpha}A^{\alpha}\nabla_{\beta}A^{\beta}+2e_{a}^{\mu}e_{b}^{\nu}n_{\alpha}A^{\alpha}\nabla_{\{\mu}A_{\nu\}}\\
\label{IsraelInTV}
&~-e_{a}^{\mu}e_{b}^{\nu}n^{\rho}\nabla_{\rho}A_{\{\mu}A_{\nu \}}-e_{a}^{\mu}e_{b}^{\nu}A_{\mu}A_{\nu}\mathcal{K}\big)\big\}_\pm=-\mathcal{S}_{ab}
\end{align}
in which $\mathcal{K}_{ab}=e^\mu _a e^\nu _b \mathcal{K}_{\mu\nu}$ with  the extrinsic curvature tensor $\mathcal{K}_{\mu\nu}$ defined by $K_{\mu \nu}=\frac{1}{2}\big( \nabla_\mu n_\nu +\nabla_\nu n_\mu \big)$. Besides, the convention $\{X\}_\pm$ denotes $\{X\}_\pm=X\big\vert_{\gamma^+_{ab}}-X\big\vert_{\gamma^-_{ab}}$. Thus, if the spacetime geometry is smooth at junction position, the energy-momentum tensor will vanish. After expanding $\eqref{IsraelInTV}$ explicitly, the following two independent equations are deduced
\begin{widetext}
	\begin{align}
	\label{Junc00com}
	&\mathcal{S}_{0}^{0}=\frac{\sqrt{1-b(r_{1})/r_{1}}}{16r_{1}^{2}}\bigg(3r_{1}^{2}a(r_{1})^{2}e^{-\Phi(r_{1})}\big(\Phi_{+}^{\prime}-\Phi_{-}^{\prime}\big)-3\big(r_{1}-b(r_{1})\big)\big(r_{1}(\Phi_{+}^{\prime}\chi_{+}^{2}-\Phi_{-}^{\prime}\chi_{-}^{2})-4(\chi_{+}^{2}-\chi_{-}^{2})\big)\bigg)\\
	\label{Juncijcom}
	&\mathcal{S}_{j}^{i}=-\frac{\sqrt{1-b(r_{1})/r_{1}}}{16r_{1}}\bigg(3r_{1}a(r_{1})^{2}e^{-\Phi(r_{1})}\big(\Phi_{+}^{\prime}-\Phi_{-}^{\prime}\big)+8r_{1}\big(\Phi_{+}^{\prime}-\Phi_{-}^{\prime}\big)-3\big(r_{1}-b(r_{1})\big)\big(\chi_{+}^{2}\Phi_{+}^{\prime}-\chi_{-}^{2}\Phi_{-}^{\prime}\big)\bigg)
	\end{align}
\end{widetext}
From the expressions $\eqref{Junc00com}-\eqref{Juncijcom}$, it is easy to see that both $\mathcal{S}^0_0$ and $\mathcal{S}^i_j$ vanish when $\Phi$ and $\chi$ are continuous in first derivatives at $r=r_1$. Thus, the spacetime geometry is smooth at the junction position.

\begin{figure}[ht]
	\begin{center}
		\includegraphics[scale=0.33]{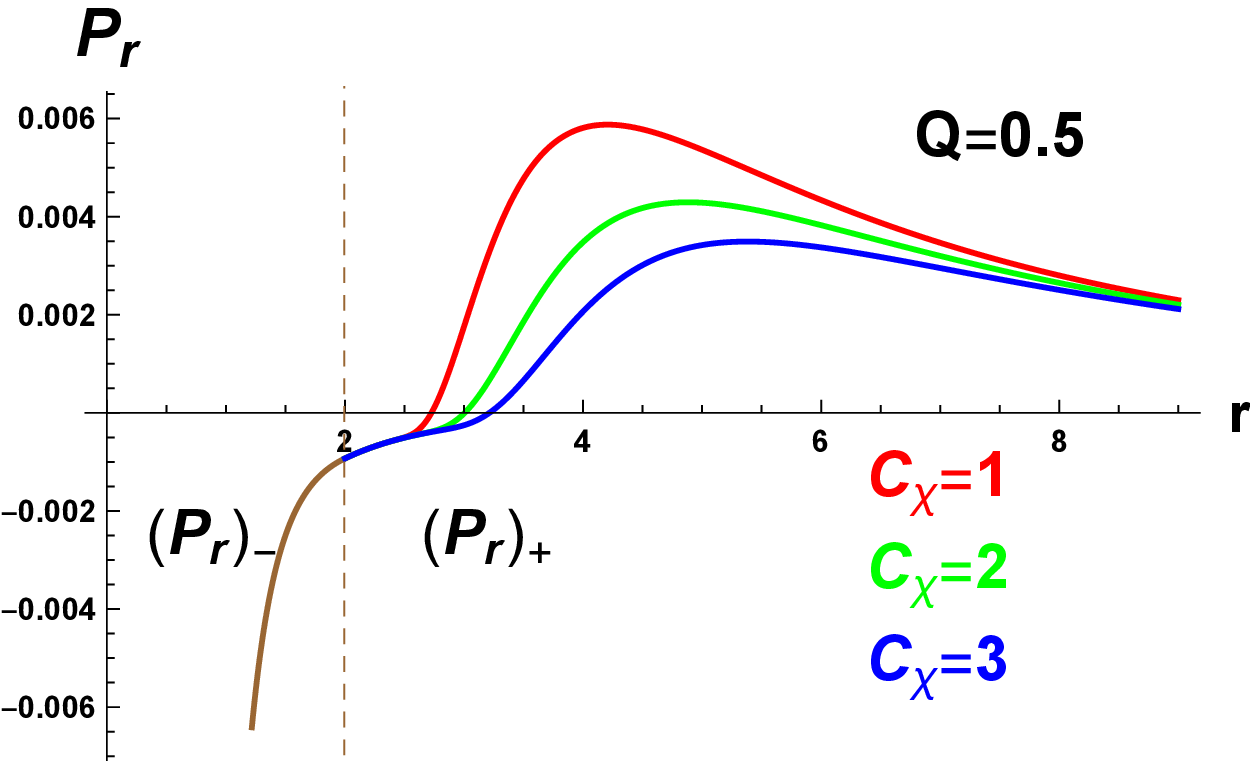}
		\includegraphics[scale=0.33]{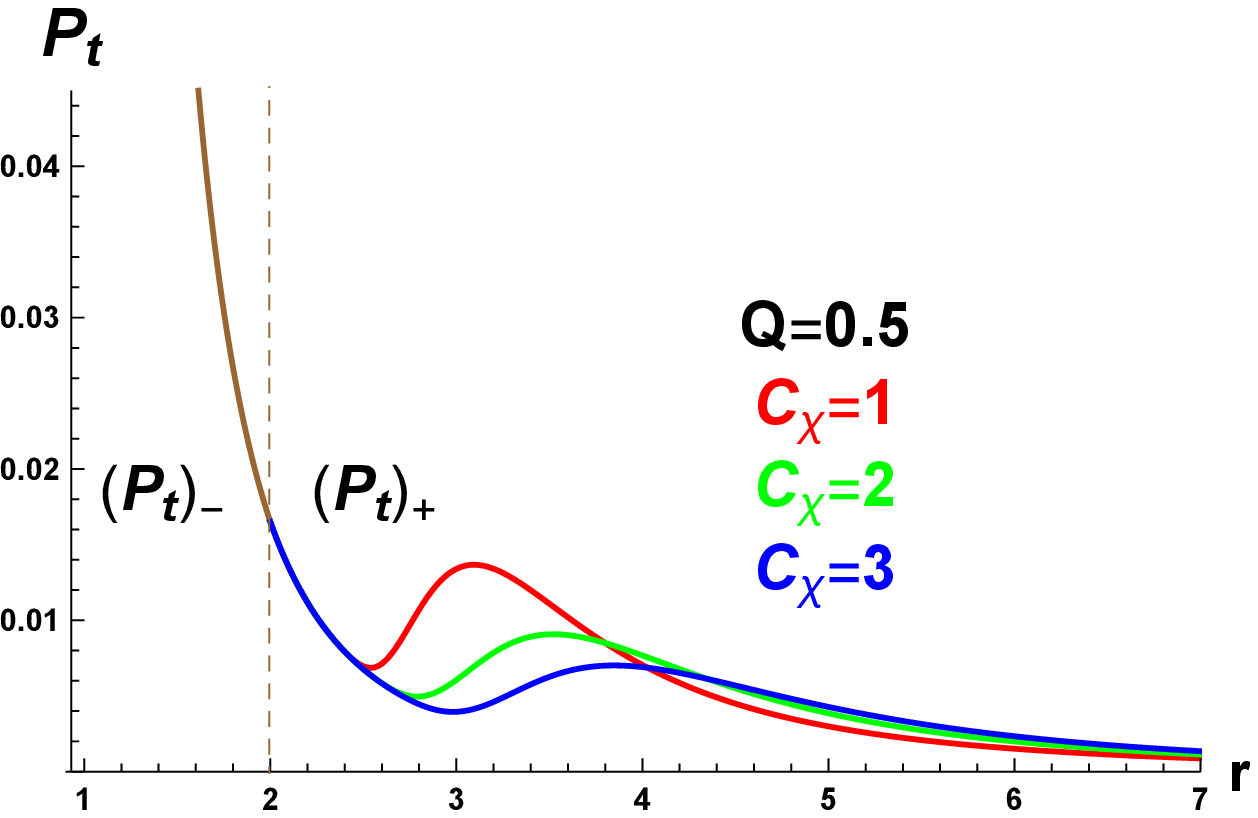}\\
		\includegraphics[scale=0.33]{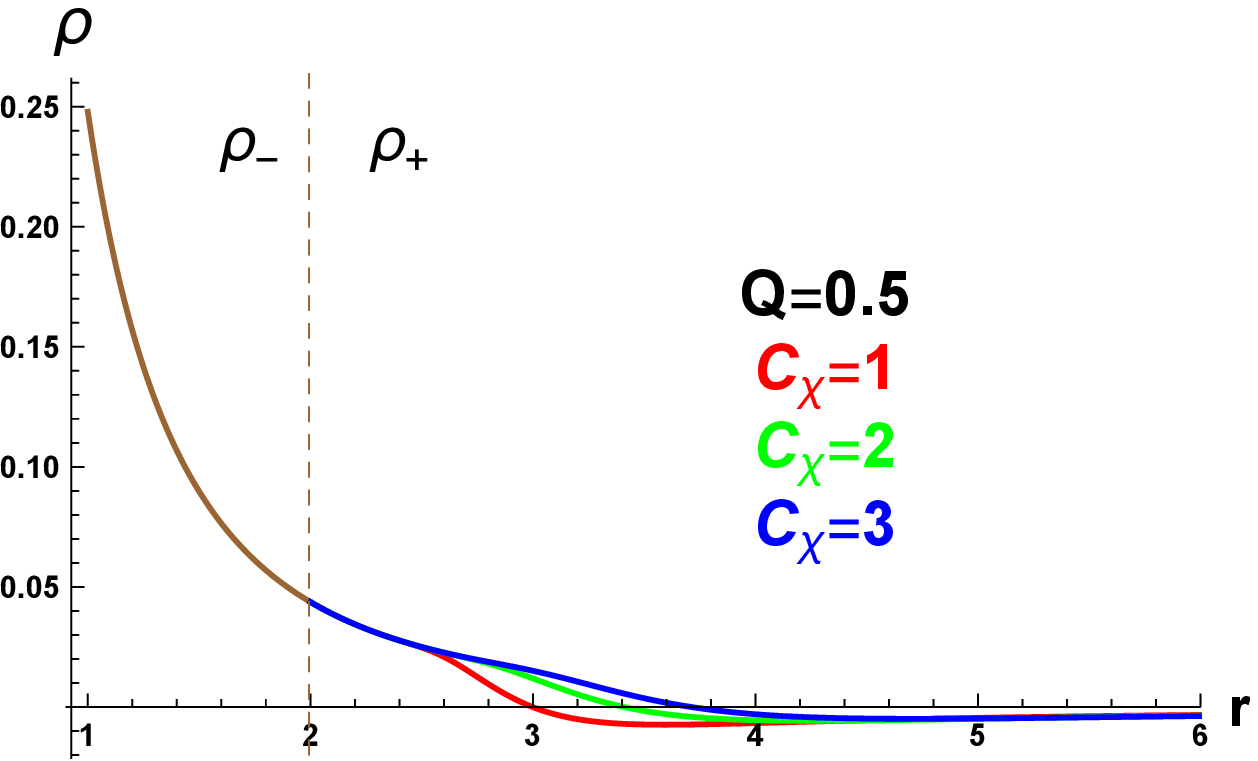}
		\includegraphics[scale=0.33]{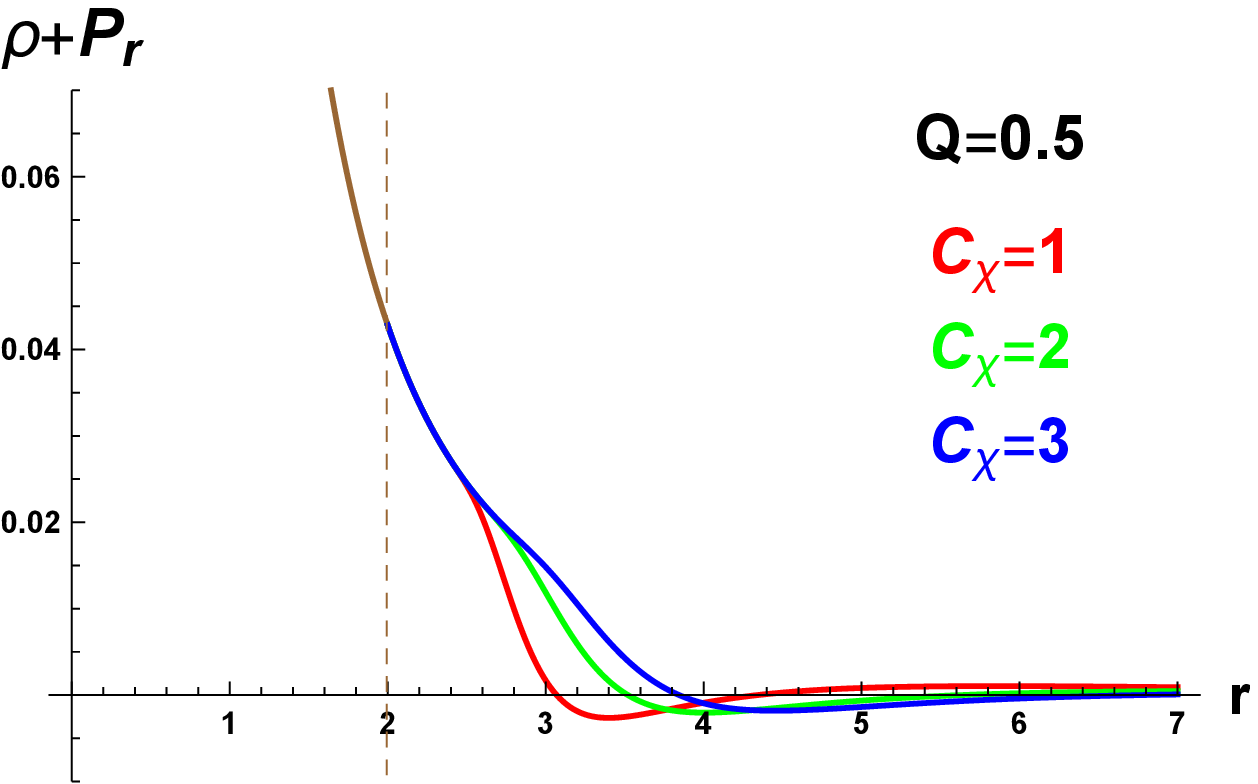}
		\caption{(color online). Plot the variation of $\rho,~P_r,~P_t,~\rho+P_r$ with respect to $r$ in small $Q$, with different $C_\chi$. Here, the $r_0,~r_1,~\alpha,~c$ are chosen as $1,~2,~\frac{1}{2},~\frac{1}{2}$ respectively. }
		\label{EMtenQs}
	\end{center}
\end{figure}

\begin{figure}[ht]
	\begin{center}
		\includegraphics[scale=0.33]{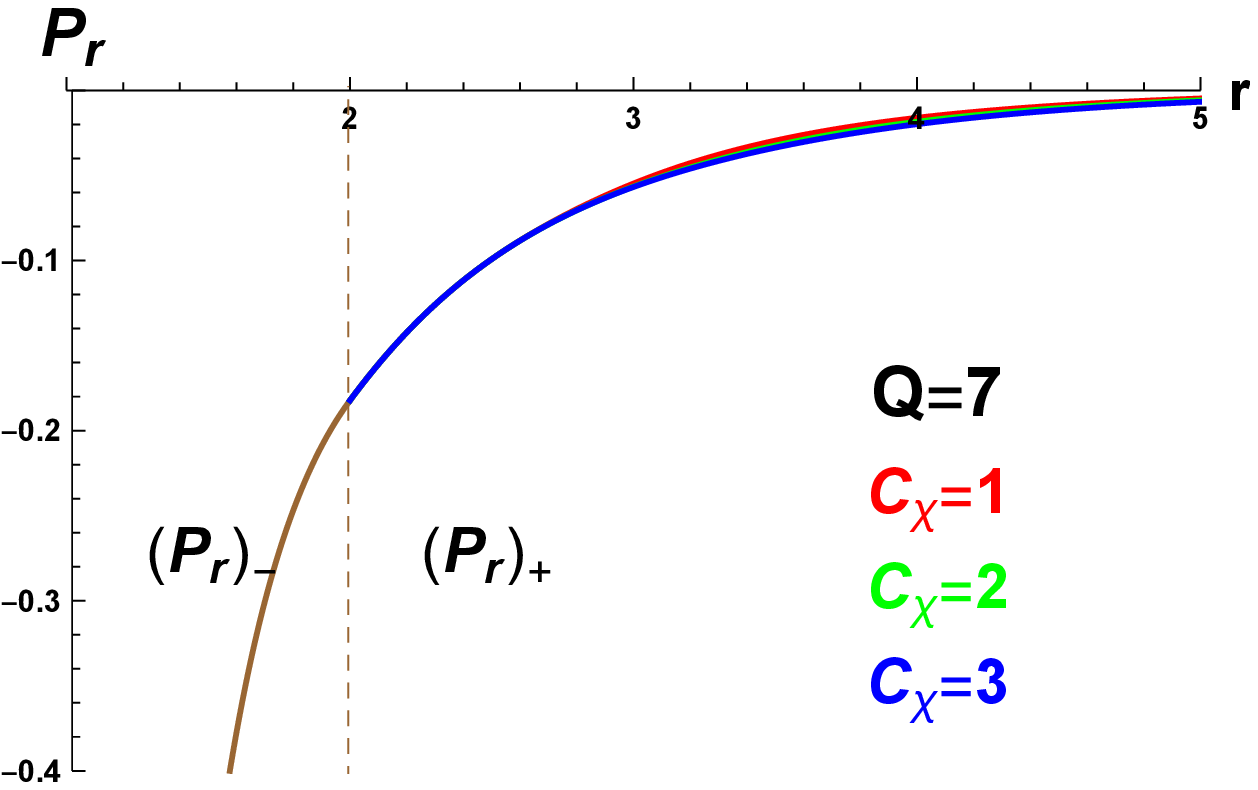}
		\includegraphics[scale=0.33]{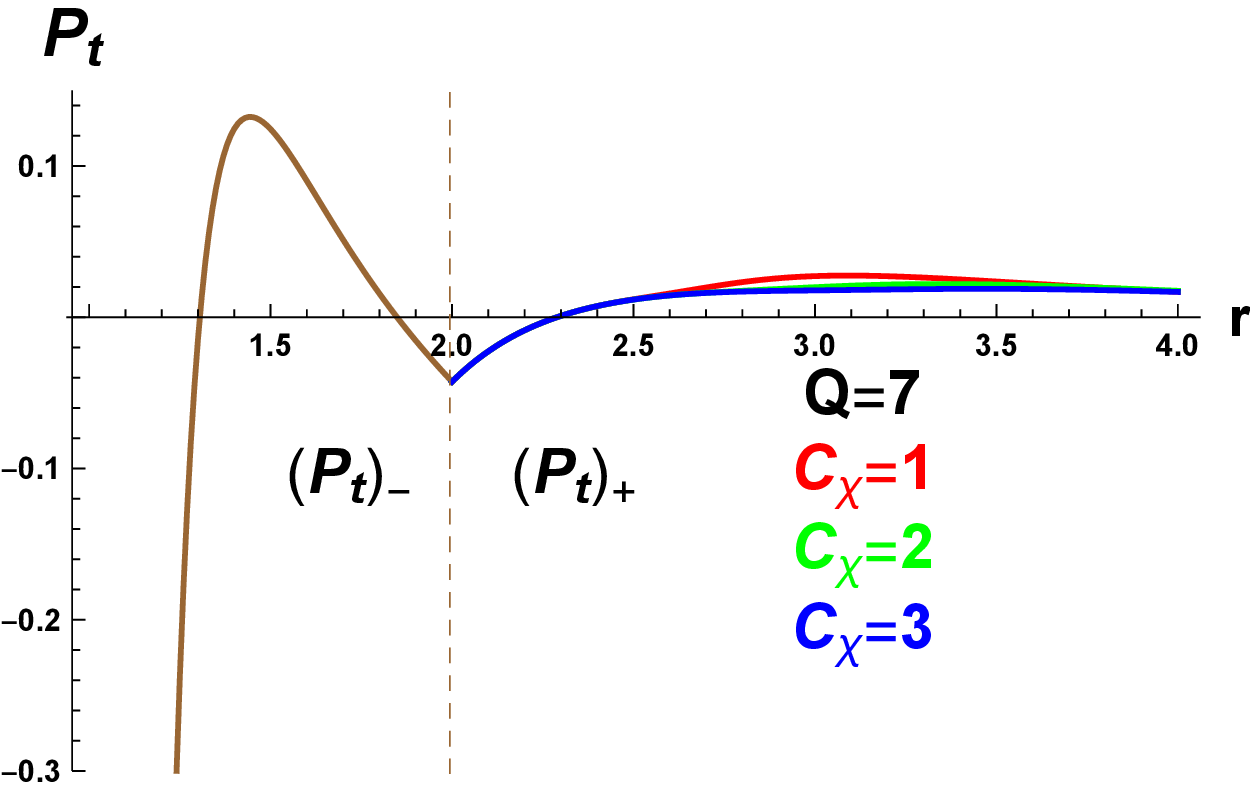}\\
		\includegraphics[scale=0.33]{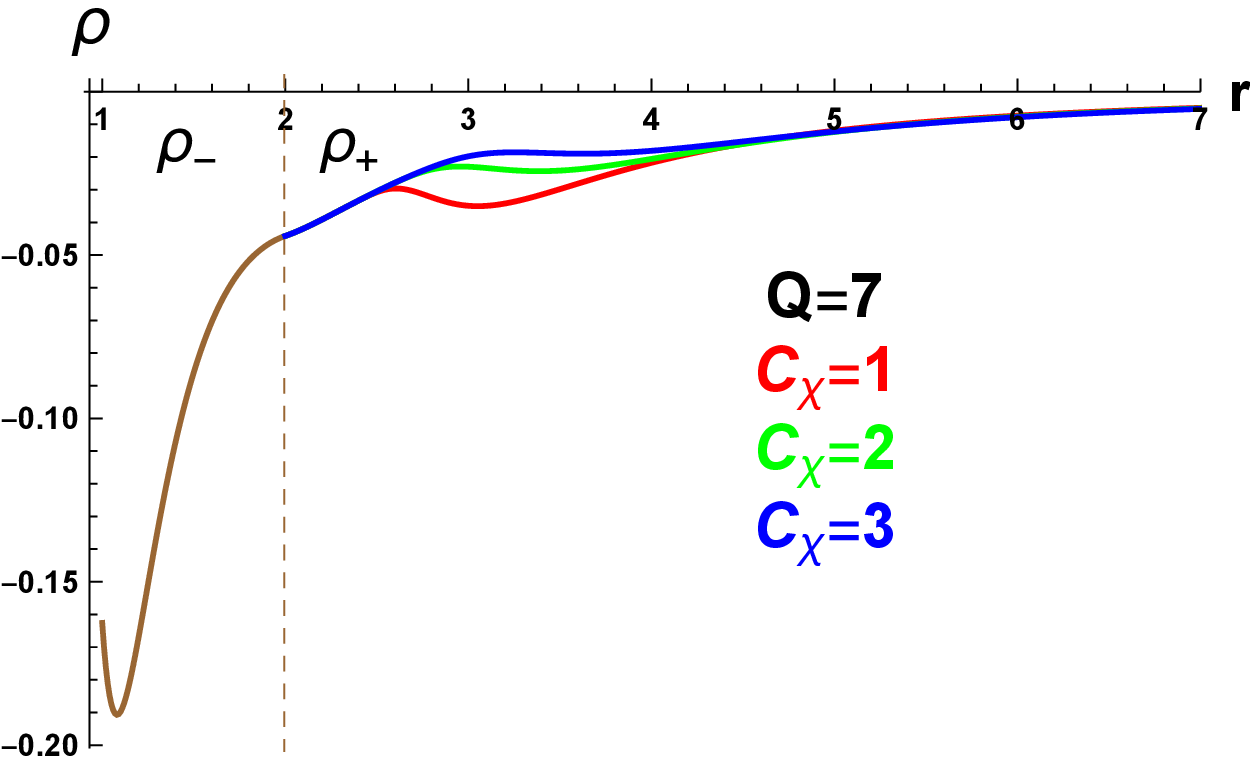}
		\includegraphics[scale=0.33]{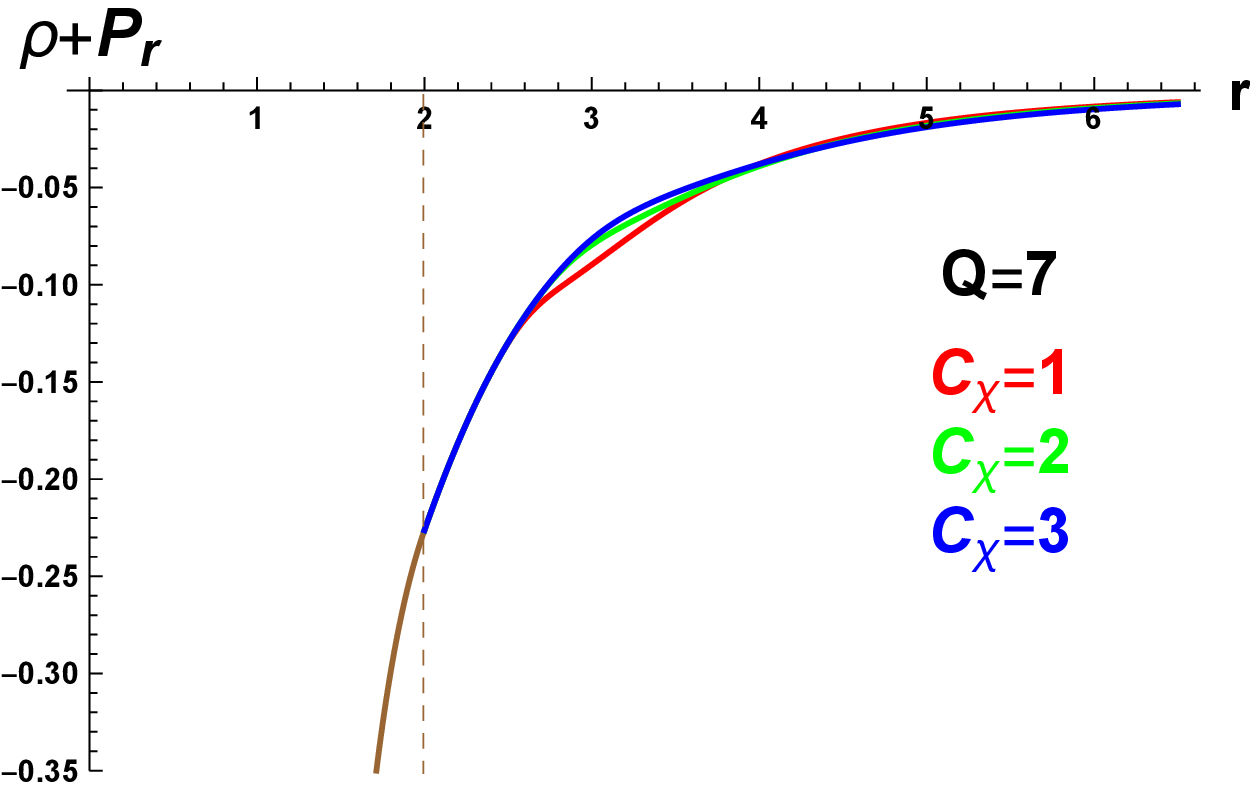}
		\caption{(color online). Plot the variation of $\rho,~P_r,~P_t,~\rho+P_r$ with respect to $r$ in large $Q$, with different $C_\chi$. Here, the $r_0,~r_1,~\alpha,~c$ are chosen as $1,~2,~\frac{1}{2},~\frac{1}{2}$ respectively.}
		\label{EMtenQL}
	\end{center}
\end{figure}

After substituting $\eqref{pisefunctionchi}$-$\eqref{pisefunctionPhiMinu}$ into the $\eqref{Eintt}$-$\eqref{Einthetatheta}$, the variation of $\rho,~P_r,~P_t,~\rho+P_r$ with respect to $r$ in some representative parameters could be shown in Fig.\ref{EMtenQs} and Fig.\ref{EMtenQL} respectively.
Thus, when $\chi(r)$ is turned on, WEC and NEC are broken in both small $Q$ and large $Q$. Similar to the case of $\chi=0$, the total amount of exotic matter in Fig.\ref{EMtenQs} and Fig.\ref{EMtenQL} could be evaluated by integral $\eqref{TotaMountExo}$, which are finite due to the fact that the value of $\rho+P_r$ is convergent at $r=r_0$.

\section{Conclusions and Discussion \label{ConDiss}}

In this paper, we have constructed an asymptotically flat Morris-Thorne wormhole in 4-dimensional spacetime, which is supported by anisotropic fluid and a vector field coupled to gravity in a non-minimal way with broken Abelian gauge symmetry. Throughout our discussion, the shape function $b(r)$ is chosen as the specific function $\eqref{SpeShape1}$. Meanwhile, the solution of $\Phi(r)$ is associated to the ansatz of vector field $A_\mu(r)$. Firstly, we suppose the vector field has the form $A_\mu dx^\mu=a(r)dt$, which implies that there exists the electrostatic potential only. Then, in order to simplify the calculations, the redshift function $\Phi(r)$ is considered as a constant value, namely a wormhole solution without tidal force. Under these conditions, as shown in Fig.\ref{EMvsQspec1}, it is easy to observe that the WEC and the NEC hold in all ranges of $r$ when $Q$ is small but are broken near the $r_0$ as the value of $Q$ increases. Besides, when the NEC and the WEC are violated, we find that the total amount of exotic matter is finite according to the volume integral quantifier $\eqref{TotaMountExo}$.

Furthermore, if the vector potential in $r$-direction is turned on, i.e. $A_\mu dx^\mu=a(r)dt+\chi(r)dr$, the redshift function will be determined by the $r$-component of extended Maxwell equations $\eqref{Vecr}$. Since this equation is invalid at the wormhole's throat $r=r_0$, in order to let $\eqref{Vecr}$ hold in all ranges of $[r_0,\infty)$,  $\chi(r)$ is assumed to possess the expression as $\eqref{pisefunctionchi}$ to keep continuity in first derivative at junction position $r=r_1$. Correspondingly, $\Phi(r)$ behaves as the piecewise functions $\eqref{pisefunctionPhi}$. In $r\geq r_1$, $\Phi(r)$ is determined by solving equation $\eqref{DeterPhi}$. Whereas, equation $\eqref{DeterPhi}$ is trivial in $r_0\leq r \leq r_1$ since $\chi(r)$ vanishes in this region. Thus, in order to keep $\Phi(r)$ finite in $r_0\leq r \leq r_1$ and continuous at junction position $r_1$, the specific function $\eqref{pisefunctionPhiMinu}$ is chosen. Besides, by evaluating the Israel junction condition, we prove that the spacetime geometry is smooth at the junction position $r_1$ if $\Phi(r)$ and $\chi(r)$ are continuous for their first derivatives. Finally, in case of $\chi\ne0$, as displayed in Fig.\ref{EMtenQs}-Fig.\ref{EMtenQL}, both the WEC and the NEC are broken whatever the value of $Q$.

For the future research, it is interesting to mention the following extended topics. In this work, we have ignored the effects of cosmological constant. Thus, it is worthwhile to construct the static, asymptotically AdS Morris-Thorne wormhole in the vector-tensor theory when $\Lambda$ is involved. Besides, as in the work \cite{Roman:1992xj}, our work could be generalized to study the Lorentzian wormholes in cosmic inflation which takes consideration of the non-minimal coupling between the vector field and the gravity with broken Abelian gauge symmetry \cite{Golovnev:2008cf, Koivisto:2008xf}.

\section{Acknowledgements}
Ai-chen Li is supported by the University of Barcelona (UB) / China Scholarship Council (CSC) joint scholarship and NSFC grant no.11875082. Xin-Fei Li is supported by the Doctor Start-up Foundation of Guangxi University of Science and Technology with Grant No. 19Z21.

\end{document}